\documentclass[prb,twocolumn,showpacs]{revtex4-1}

\usepackage{amsmath}
\usepackage{graphicx}
\usepackage{bm}

\begin{document}

\title{Crossover between two different magnetization reversal modes \\
in arrays of iron oxide nanotubes}
\author{J. Escrig$^{1,2}$}
\author{J. Bachmann$^{1,3}$}
\author{J. Jing$^1$}
\author{M. Daub$^1$}
\author{D. Altbir$^2$}
\author{K. Nielsch$^{1,3}$}

\affiliation{$^{1}$Max Planck Institute of Microstructure Physics, Am Weinberg 2, 06120
Halle, Germany\\
$^{2}$Departamento de F\'{\i}sica, Universidad de Santiago de Chile, USACH,
Av. Ecuador 3493, Santiago, Chile\\
$^{3}$Institute of Applied Physics, University of Hamburg, Jungiusstr. 11,
20355 Hamburg, Germany}
\date{\today}

\begin{abstract}
The magnetization reversal in ordered arrays of iron oxide nanotubes of 50
nm outer diameter grown by atomic layer deposition is investigated
theoretically as a function of the tube wall thickness, $d_{w}$. In thin
tubes ($d_{w}<13$ nm) the reversal of magnetization is achieved by the
propagation of a vortex domain boundary, while in thick tubes ($d_{w}>13$
nm) the reversal is driven by the propagation of a transverse domain
boundary. Magnetostatic interactions between the tubes are responsible for a
decrease of the coercive field in the array. Our calculations are in
agreement with recently reported experimental results. We predict that the
crossover between the vortex and transverse modes of magnetization reversal
is a general phenomenon on the length scale considered.
\end{abstract}

\maketitle

\section{Introduction}

Magnetic nanoparticles, particles of nanometer size made from magnetic
materials, have attracted increasing interest among researchers of various
fields due to their promising applications in hard disk drives, magnetic
random access memory, and other spintronic devices. \cite%
{SMW+00,KDA+98,CKA+99,WAB+01,GBH+02} Besides, these magnetic nanoparticles
can be used for potential biomedical applications, such as magnetic
resonance imaging (the nanoparticles can be used to trace bioanalytes in the
body), cell and DNA separation, and drug delivery. \cite{ET03} To apply
nanoparticles in various potential devices and architectures, it is very
important to control the size and shape and to keep the thermal and chemical
stability of the nanoparticles. \cite{PKA+01}

The properties of virtually all magnetic materials are controlled by domains
- extended regions where the spins of individual electrons are tightly
locked together and point in the same direction. Where two domains meet, a
domain wall forms. Measurements on elongated magnetic nanostructures \cite%
{WDM+96} highlighted the importance of nucleation and propagation of a
magnetic boundary, or domain wall, between opposing magnetic domains in the
magnetization reversal process. Domain-wall propagation in confined
structures is of basic interest. \cite{AAX+03, THJ+06} For instance, by
equating the direction of a domain's magnetization with a binary 0 or 1, a
domain wall also becomes a mobile edge between data bits: the
pseudo-one-dimensional structure can thus be thought of as a physical means
of transporting information in magnetic form. This is an appealing
development, because computers currently record information onto their hard
disk in magnetic form. \cite{Cowburn07}

The trusty sphere remains the preferred shape for nanoparticles but this
geometry leaves only one surface for modification, complicating the
generation of multifunctional particles. Thus, a technology that could
modify differentially the inner and outer surface would be highly desirable. 
\cite{Eisenstein05} On the one hand, over the past years there has been a
surge in research on nanocrystals with core/shell architectures. Although
extensive studies have been conducted on the preparation of
core/shell-structured nanoparticles, the fabrication and characterization of
bimetallic core/shell particles with a total size of less than 10 nm and
with a monolayer metal shell remain challenging tasks. \cite{CYD+08} On the
other hand, since the discovery of carbon nanotubes by Iijima in 1991, \cite%
{Iijima91} intense attention has been paid to hollow tubular nanostructures
because of their particular significance for prospective applications. In
2002 Mitchell \textit{et al}. \cite{MLT+02} used silica nanotubes offering
two easy-to-modify surfaces. More recently, magnetic nanotubes have been
grown \cite{SRH+05,NCM+05,Wang05,FMY+06} that may be suitable for
applications in biotechnology, where magnetic nanostructures with low
density, which can float in solutions, become much more useful for in vivo
applications. \cite{Eisenstein05} In this way tiny magnetic tubes could
provide an unconventional solution to several research problems, and a
useful vehicle for imaging and drug delivery applications.

Although the magnetic behavior of nanowires has been intensely investigated,
tubes have received less attention, in spite of the additional degree of
freedom they present; not only can the length, $L$, and radius, $R$, be
varied, but also the thickness of the wall, $d_{w}$. Changes in thickness
are expected to strongly affect the mechanism of magnetization reversal, and
thereby, the overall magnetic behavior. \cite{ELA+07,SSS+04} However,
systematic experimental studies on this aspect were lacking for a long time,
mostly due to the difficulty in preparing ordered nanotube samples of very
well-defined and tunable geometric parameters.

We recently reported the synthesis and magnetic characterization of a series
of Fe$_{3}$O$_{4}$ nanotube arrays (length $L=3$ $\mu$m, radius $R=25$ nm,
center-to-center distance $D=105$ nm, and wall thickness $2.5$ nm $<d_{w}<22$
nm), prepared by atomic layer deposition (ALD) in a porous alumina matrix. 
\cite{BJK+07} In this series, the magnetic response of the array,
characterized by the coercive field $H_{c}$ and the relative remanence, \cite%
{remanence} vary strongly and non-monotonically as a function of $d_{w}$.
For the thinner tubes, $H_{c}$ is enhanced by increasing $d_{w}$, until $%
d_{w}=13$ nm, where it presents a maximum of about $780$ Oe. For further
increases of $d_{w}$, the coercive field decreases. A quantitatively similar
behavior was also observed in Ni$_{80}$Fe$_{20}$ nanowire arrays, \cite%
{GSA+05} a different system in terms of geometry, material and preparation
techniques.

This convergence of experimental observations may reflect an underlying
general phenomenon. Therefore, this paper focuses on the investigation of
the non-monotonic behavior of the coercive field in ferromagnetic nanotube
arrays, a question that has remained unexplained until now. We start by
modelling the magnetization reversal and calculate $H_{c}$ for the system
reported experimentally, \cite{BJK+07} then generalize our conclusions and
quantitatively predict trends for other geometries and materials.

\section{Experimental methods}

Our approach to the preparation of magnetic nanotubes of well-controlled and
tunable geometric parameters and arranged in hexagonally ordered, parallel
arrays is based on the combination of two complementary aspects, namely 
\textit{(i)} the utilization of self-ordered anodic alumina (AA) as a porous
template, and \textit{(ii)} the conformal coating of its cylindrical pores
with thin oxide films by atomic layer deposition (ALD).

\textit{Anodic alumina} is obtained from the electrochemical oxidation of
aluminum metal under high voltage (usually 20 to 200 V) in aqueous acidic
solutions. \cite{MF95,NCS+02} Under certain proper sets of experimental
conditions (nature and concentration of the acid, temperature and applied
voltage), the electrochemically generated layer of alumina displays a
self-ordered porous structure. Cylindrical pores of homogeneous diameter are
thus obtained, with their long axis perpendicular to the plane of the
alumina layer and ordered in a close-packed hexagonal arrangement. With our
method, anodization of Al in $0.3$ M oxalic acid under $40$ V at $8^{o}$C
yields pores of $\sim 50$ nm outer diameter and with a center-to-center
distance of $\sim 105$ nm (an approach which we will call \textit{Method A}%
); anodization in $1\%$ phosphoric acid under $195$ V at $0^{o}$C yields
pores of $\sim 160$ nm outer diameter and with a center-to-center distance
of $\sim 460$ nm (\textit{Method B}).

\textit{Atomic layer deposition} is a self-limited gas-solid chemical
reaction. \cite{Puurunen05} Two thermally stable gaseous precursors are
pulsed alternatively into the reaction chamber, whereby direct contact of
both precursors in the gas phase is prevented. Because each precursor
specifically reacts with chemical functional groups present on the surface
of the substrate (as opposed to non-specific thermal decomposition), one
monolayer of precursor adsorbs onto the surface during each pulse despite an
excess of it in the gas phase. This peculiarity of ALD makes it suitable for
coating substrates of complex geometry (in particular, highly porous ones)
conformally and with outstanding thickness control. \cite{LRG03} We have
successfully used ALD to create Fe$_{2}$O$_{3}$ nanotubes in porous anodic
alumina templates from two different chemical reactions with similar
results. In \textit{Method I}, oxidation of ferrocene (also called
bis(cyclopentadienyl)iron, usually abbreviated Cp$_{2}$Fe) with an
ozone/dioxygen (O$_{3}$ / O$_{2}$) mixture at 200$^{o}$C yields a growth
rate of $\sim 0.2$ $\mathring{A}$ per cycle. \textit{Method II} consists in
the reaction of the dimeric iron(III) \textit{tert}-butoxide, Fe$_{2}$(O$%
^{t} $Bu)$_{6}$, with water at 140$^{o}$C, with $\sim 0.25$ $\mathring{A}$
deposited per cycle. \cite{BJK+07} Both methods yield a wall thickness
distribution within each sample below 10$\%$.

The results are shown in Figs. 1 and 2. Smooth tubes of $50$ or $160$ nm
outer diameters can be obtained, with aspect ratios on the order of $100$.
The thickness of the wall can be accurately controlled between $1$ and $50$
nm. Subsequent reduction of the Fe$_{2}$O$_{3}$ material by H$_{2}$ at $%
400^{o}$C results in the formation of the strongly magnetic phase Fe$_{3}$O$%
_{4}$, a transformation verified by X-ray photoelectron spectroscopy and
accompanied by the expected color change from yellow, orange or brown
(depending on the thickness) to black. The structural quality of the tubes
is unaffected by reduction, \cite{BJK+07} a consequence of the very small
volume contraction caused by it. Our approach allowed us to systematically
investigate the influence of structure on magnetism in a series of samples
of Fe$_{3}$O$_{4}$ nanotube arrays prepared according to \textit{Methods A}
and \textit{II} and in which the wall thickness, $d_{w}$, varies while all
other geometric parameters are maintained constant.

\begin{figure}[h]
\begin{center}
\includegraphics[width=8cm]{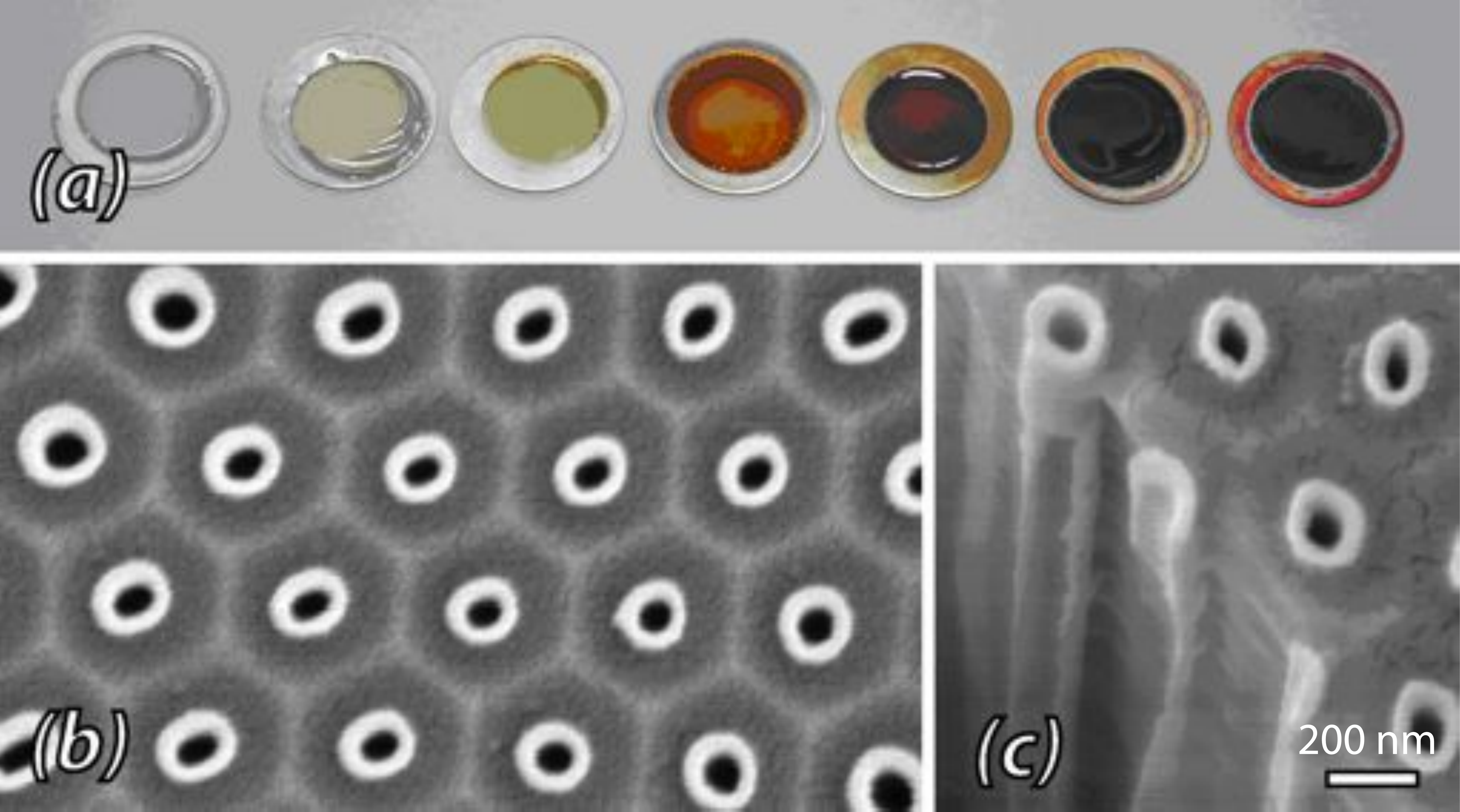}
\end{center}
\caption{(Color online) Aspects of Fe$_{2}$O$_{3}$ nanotubes grown by atomic
layer deposition (ALD) in a porous anodic alumina matrix. \textit{(a)}
Macroscopic view of the samples, consisting of a circular porous anodic
alumina (AA) membrane containing the embedded tubes, surrounded by an outer
circle of Al metal of 2 cm outer diameter; the tubes in the samples from
left to right have walls of increasing thickness, approximately 0, 1, 2, 4,
8, 12, and 16 nm, respectively (AA obtained according to \textit{Method A},
ALD performed by \textit{Method I}). \textit{(b, c)} Scanning electron
micrographs (SEM) of tubes embedded in the porous alumina matrix, observed
in top view and at an angle at a break in the sample, respectively; the
scale bar represents 200 nm (AA obtained according to \textit{Method B}, ALD
performed by \textit{Method II}).}
\end{figure}

\begin{figure}[h]
\begin{center}
\includegraphics[width=5cm]{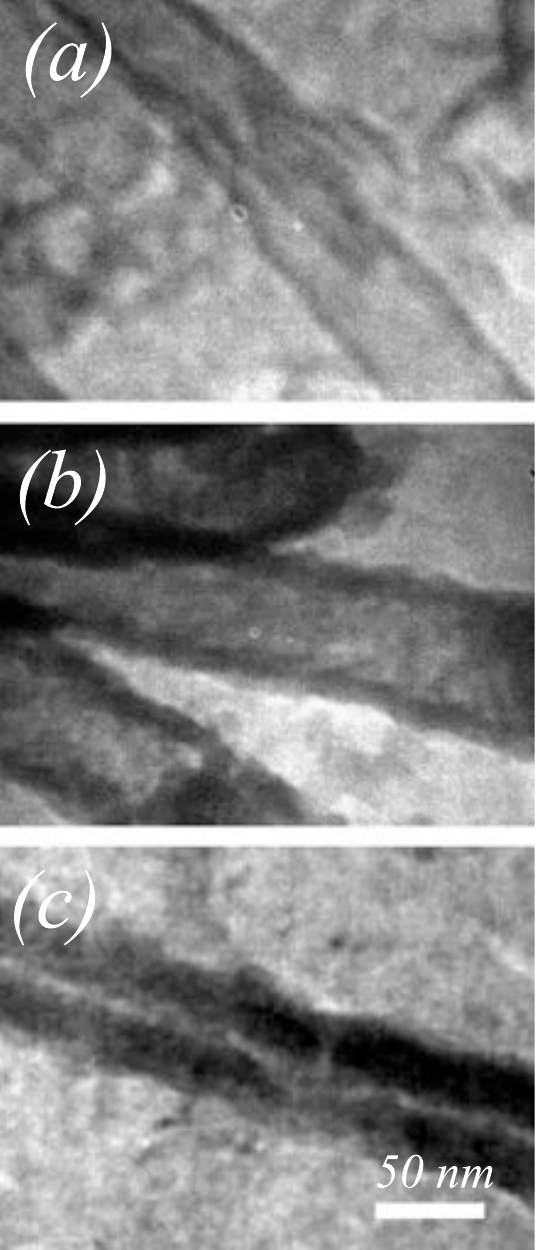}
\end{center}
\caption{Transmission electron micrographs (TEM) of isolated tubes from the
alumina matrix, with walls of increasing thicknesses, approximately of 1%
\textit{(a)}, 5 \textit{(b)}, and 13 \textit{(c)} nm, respectively; the
scale bar represents 50 nm (AA obtained according to \textit{Method A}, ALD
performed by \textit{Method II}).}
\end{figure}

In a series of Fe$_{3}$O$_{4}$ nanotube arrays of varying wall thickness $%
d_{w}$ (all other geometric parameters being kept constant), investigated by
SQUID magnetometry, we observed a significant dependence of the coercivity
and remanence upon the geometry. In particular, the coercive field $H_{c}$
can be tuned between 0 and 800 Oe (0 and 80 mT) approximately by properly
adjusting $d_{w}$. Most curiously, the dependence of $H_{c}$ on $d_{w}$ is
not monotonic - $H_{c}$ reaches its maximum at $d_{w}\approx 13$ nm and then
decreases for further increases of the wall thickness (Figure 6b). We
interpret this observation as arising from the coexistence of two distinct
magnetization reversal modes in our system. Which of the two prevails in a
given sample is uniquely determined by the geometric parameters of the tube
array. Thus, the cusp in the $H_{c}\left( d_{w}\right) $ curve corresponds
to the crossover between the two modes of magnetization reversal. The
following paragraphs detail the theoretical treatment of the two modes.

\section{Two magnetization reversal modes}

For isolated magnetic nanotubes, the magnetization reversal, that is, the
change of the magnetization from one of its energy minima ($\mathbf{M}=M_{0}$
$\mathbf{\hat{z}}$) to the other ($\mathbf{M}=-M_{0}$ $\mathbf{\hat{z}}$),
can occur by one of only two idealized mechanisms, the Vortex mode (V),
whereby spins in rotation remain tangent to the tube wall, or the Transverse
mode (T), in which a net magnetization component in the ($x$, $y$) plane
appears. \cite{LAE+07} In both cases, a domain boundary appears at one end
of the tube and propagates towards the other, as illustrated in Fig. 3.
Starting from the equations presented by Landeros \textit{et al}, \cite%
{LAE+07} we can calculate the zero-field energy barrier as well as the width
of the domain boundary for each reversal mode as a function of the tube
thickness, $d_{w}$. Figure 4a, 4b, and 4c present our results for Fe$_{3}$O$%
_{4}$ nanotubes using $M_{0}=4.8\times 10^{5}$ A/m$^{3}$ and the stiffness
constant $A=10^{-11}$ J/m. \cite{Ohandley00} Figure 4a witnesses a crossover
at $d_{w}\approx 20$ nm, showing that the V mode is more stable for thinner
tubes, whereas thicker tube walls favor the T mode. This result can be
qualitatively explained as follows. A very thin tube should behave as a
(rolled-up) thin film, in which the magnetic moments always tend to remain
within the plane of the film. Conversely, tubes of large wall thicknesses
approach the case of wires: surface effects are less crucial, but
interactions between diametrically opposed regions become more important.

\begin{figure}[h]
\begin{center}
\includegraphics[width=7cm]{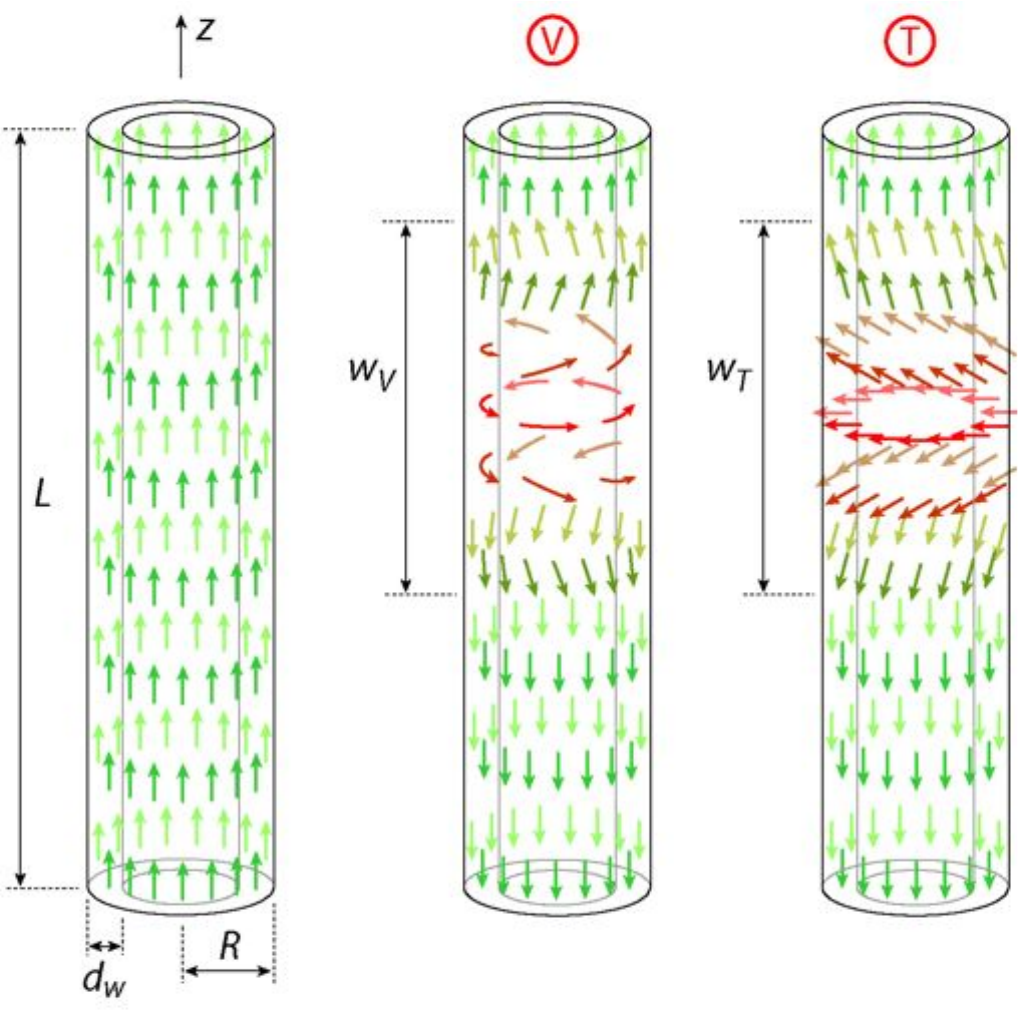}
\end{center}
\caption{(Color Online) Geometric parameters and magnetization reversal
modes in nanotubes. Arrows represent the orientation of magnetic moments
within the tube. Left: A magnetically saturated tube ($\mathbf{M}=M_{0}$ $%
\mathbf{\hat{z}}$), with its geometric parameters, length $L$, radius $R$,
wall thickness $d_{w}$. Center: A tube during the magnetization reversal
from $M_{0}$ $\mathbf{\hat{z}}$ to $-M_{0}$ $\mathbf{\hat{z}}$ via a vortex
mode, V; the domain boundary, of thickness $w_{V}$, migrates upwards. Right:
The equivalent situation by means of a transverse reversal mode, T, with a
domain boundary of thickness $w_{T}$.}
\end{figure}

\begin{figure}[h]
\begin{center}
\includegraphics[width=7cm]{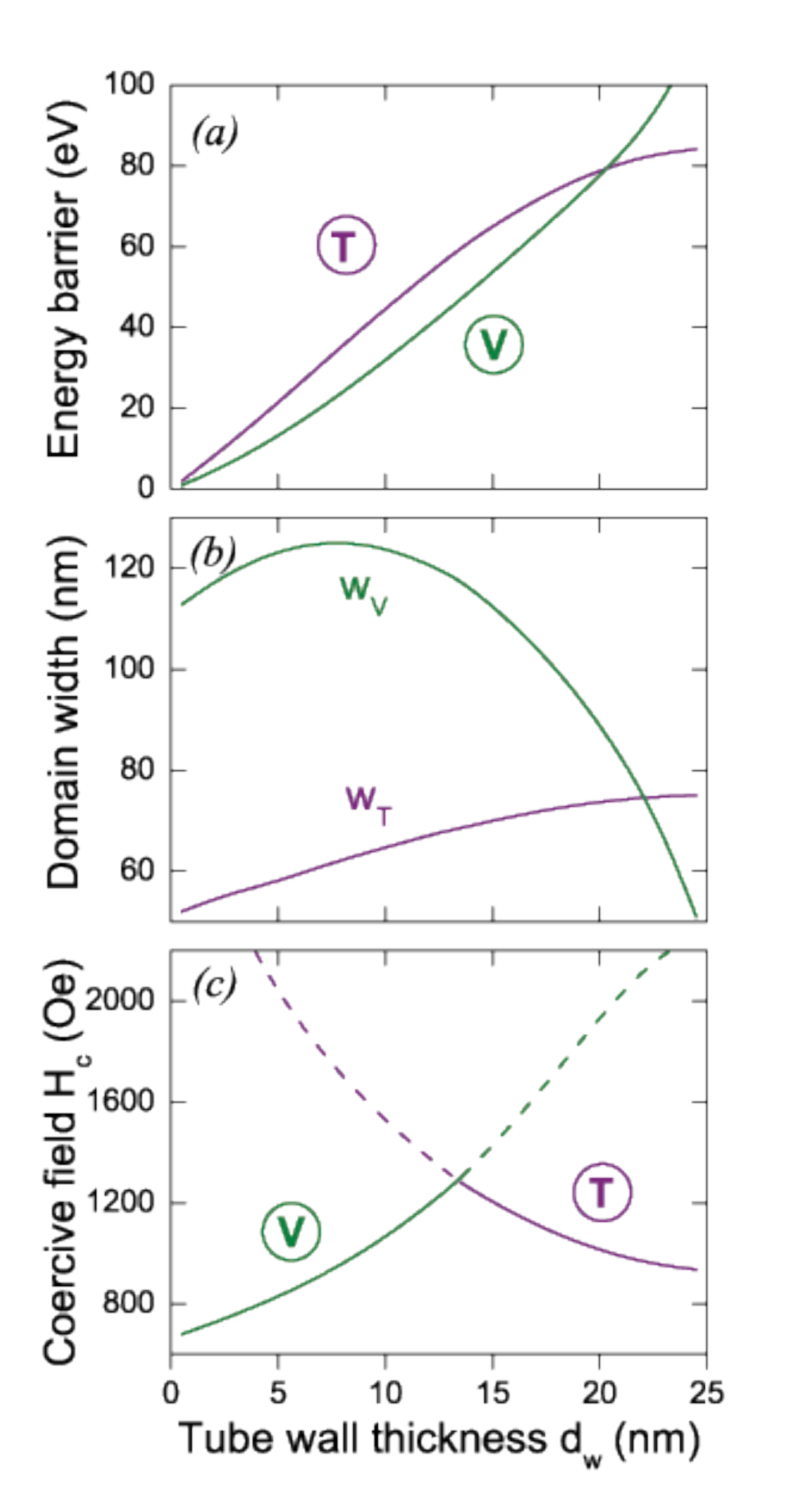}
\end{center}
\caption{(Color online) Energy barrier \textit{(a)} and domain boundary
width \textit{(b)} as a function of tube wall thickness for the V (green)
and T (purple) modes of magnetization reversal. \textit{(c)} Coercive fields
obtained from Eqs. (1) and (2) for the V and T modes, (respectively shown in
green and purple) for tube wall thicknesses varying between 0.5 and 24.5 nm.
The dashed lines of the curves have no physical meaning. We have used $R=25$
nm}
\end{figure}

The presence of a crossover in Fig. 4a allows us to expect a transition from
the V to the T reversal mode with increasing values of $d_{w}$. However, the
curves cannot give the coercive field values directly because energies
represent the difference between a completely saturated state and one with a
domain boundary in the middle of the tube. Magnetization reversal, however,
is initiated with a domain boundary at one end of the tube, a configuration
that corresponds to a lower magnetic energy. From Fig. 4b we observe changes
in boundary widths between 50 and 120 nm as a function of $d_{w}$. A
crossover is found, corresponding to the one that appears between the energy
curves. We shall now proceed to calculate the switching field of an isolated
magnetic nanotube assuming that the magnetization reversal is driven by
means of one of the two previously presented modes. \cite{coercivity}

\subsection{Coercive fields}

For the T mode, the coercive field, $H_{n}^{T}$, can be approximated by an
adapted Stoner-Wohlfarth model \cite{SW48} in which the length of the
coherent rotation is replaced by the width of the domain boundary, $w_{T}$
(see Fig. 4b). Following this approach, 
\begin{equation}
\frac{H_{n}^{T}}{M_{0}}=\frac{2K\left( w_{T}\right) }{\mu _{0}M_{0}^{2}}\ ,
\label{1}
\end{equation}%
where $K\left( l\right) =\frac{1}{4}\mu _{0}M_{0}^{2}\left( 1-3N_{z}\left(
l\right) \right) $ and $N_{z}\left( l\right) $ corresponds to the
demagnetizing factor along $z$, given by $N_{z}\left( l\right) =\frac{2R}{%
\left( 1-\beta ^{2}\right) l}\int_{0}^{\infty }\frac{dq}{q^{2}}\left(
J_{1}\left( q\right) -\beta J_{1}\left( q\beta \right) \right) ^{2}\left(
1-e^{-q\frac{l}{R}}\right) $, with $\beta =1-\frac{d_{w}}{R}$.

For the V mode we use an expression for the nucleation field obtained by
Chang \textit{et al}. \cite{CLY94} When an external field with magnitude
equal to the nucleation field is applied opposite to the magnetization of
the tube, infinitesimal deviations from the initially saturated state along
the tube axis appear. The form of these deviations is determined by the
solution of a linearized Brown's equation. \cite{AS58} Furthermore, it has
been shown numerically that the solution of this Brown's equation is not a
stable solution to the full nonlinear equation at applied fields larger than
the nucleation field, and then the only possible stable states are those
with uniform alignment along the axis. \cite{BB92,Brown58} Thus, the
magnetization is assumed to reverse completely at the nucleation field. For
an infinite tube, the nucleation field for the V mode, $H_{n}^{V}$, is given
by 
\begin{equation}
\frac{H_{n}^{V}}{M_{0}}=\alpha \left( \beta \right) \frac{L_{ex}^{2}}{R^{2}}%
\ ,  \label{2}
\end{equation}%
with $L_{ex}=\sqrt{2A/\mu _{0}M_{0}^{2}}$ and $\alpha \left( \beta \right)
\equiv q^{2}$, where $q$ satisfies the condition%
\begin{equation}
\frac{qJ_{0}\left( q\right) -J_{1}\left( q\right) }{qY_{0}\left( q\right)
-Y_{1}\left( q\right) }=\frac{\beta qJ_{0}\left( \beta q\right) -J_{1}\left(
\beta q\right) }{\beta qY_{0}\left( \beta q\right) -Y_{1}\left( \beta
q\right) }\ .  \label{3}
\end{equation}%
Here $J_{p}\left( z\right) $ and $Y_{p}\left( z\right) $ are Bessel
functions of the first and second kind, respectively. Equation (3) has an
infinite number of solutions, and the physically correct solution is the
smallest one.

Figure 4c illustrates the coercive field of an isolated tube with $d_{w}$
varying from $0.5$ to $24.5$ nm. We can observe a crossing of the two curves
at $d_{w}=13$ nm approximately, corresponding to a magnetization reversal
for which both V and T mechanisms are possible at the same coercive field.
At other given values of $d_{w}$, the system will reverse its magnetization
by whichever mode opens an energetically accessible route first, that is, by
the mode that offers the lowest coercivity. Therefore, the curve of
coercivity vs. $d_{w}$ is the solid one, and the dotted sections of curves
have no physical meaning. Thus, our Fe$_{3}$O$_{4}$ tubes will reverse their
magnetization by the V mode for $d_{w}<13$ nm, and by the T mode for $%
d_{w}>13$ nm. Our calculations for an isolated tube reproduce the
non-monotonic behavior of the coercive field as a function of the wall
thickness experimentally observed, with a transition between two different
modes causing a cusp at $d_{w}^{V-T}=13$ nm (with $d_{w}^{V-T}$ the
thickness at which the transition occurs). However, the absolute values
computed for the coercivity are greater than the experimental data.

\subsection{Effect of the stray field}

We ascribe such difference between calculations and experimental results to
the interaction of each tube with the stray fields produced by the array -
an effective antiferromagnetic coupling between neighboring tubes, which
reduces the coercive field (as previously demonstrated in the case of
nanowires; see Fig. 6a). \cite{Hertel01,VPH+04,BAA+06,ELP+08} In these
interacting systems, the process of magnetization reversal can be viewed as
the overcoming of a single energy barrier, $\Delta E$. In an array with all
the nanotubes initially magnetized in the same direction, the magnetostatic
interaction between neighboring tubes favours the magnetization reversal of
some of them. A reversing field aligned opposite to the magnetization
direction lowers the energy barrier, thereby increasing the probability of
switching. The dependence of the applied field on the energy barrier is
often described \cite{Sharrock94} by the expression%
\begin{equation*}
\Delta E=U\left( 1-\frac{H}{H_{0}}\right) ^{2}\ ,
\end{equation*}%
where $H$ is the applied field, and $H_{n}$ denotes the intrinsic coercivity
of an isolated wire. For single-domain particles having a uniaxial shape
anisotropy, the energy barrier at zero applied field, $U$, is just the
energy required to switch by coherent rotation, $K\left( L\right) $. If we
assume that the switching field $H_{s}$ is equal to $H_{c}$, then 
\begin{equation}
H_{c}=H_{n}^{i}-H_{int},
\end{equation}%
where $H_{n}^{i}$ denotes the intrinsic coercivity $H_{n}^{V}$ or $H_{n}^{T}$
of an isolated tube, and $H_{int}$ corresponds to the stray field induced
within the array given by%
\begin{equation}
\frac{H_{int}}{M_{0}}=\frac{2K\left( L\right) }{\mu _{0}M_{0}^{2}}\left( 
\frac{\varepsilon \left\vert \tilde{E}_{int}\left( s\right) \right\vert }{%
K\left( L\right) }\right) ^{1/2}\ ,  \label{5}
\end{equation}%
In the previous equation we have assumed that the reversal of individual
nanotubes produces a decrease of the magnetostatic energy $E_{int}$ that
equals the magnetic anisotropy barrier $\Delta E$. Besides, $\varepsilon $
is an adjustable parameter that depends on the distribution of magnetic
tubes in space and on the long-distance correlation among the tubes. The
value of $\varepsilon $ can not be obtained from first principles, although
values between unity and some tens could be a reasonable estimate for this
quantity. \cite{ACT+01} Besides, $\tilde{E}_{int}\left( s\right) $ is the
magnetostatic interaction between two nanotubes separated by a distance $s$.
Such interaction can be calculated by considering each tube homogeneously
magnetized and is given by%
\begin{multline*}
\tilde{E}_{int}\left( s\right) \equiv \frac{E_{int}\left( s\right) }{V}=%
\frac{2\mu _{0}M_{0}^{2}R}{\left( 1-\beta ^{2}\right) L} \\
\cdot \int_{0}^{\infty }\frac{dq}{q^{2}}J_{0}\left( q\frac{s}{R}\right)
\left( J_{1}\left( q\right) -\beta J_{1}\left( q\beta \right) \right)
^{2}\left( 1-e^{-q\frac{L}{R}}\right) .
\end{multline*}

The resulting curve $H_{int}(d_{w})$ $(\varepsilon=20)$ is illustrated in
the top panel of Fig. 6a. The stray fields produced by an array of nanotubes
are significant for the experimentally investigated tubes, being on the
order of $350$ Oe for $d_{w}=5$ nm to $580$ Oe by $d_{w}=21$ nm.

\subsection{Results}

The hysteresis loop in normalized axis of a sample of $L=3$ $\mu$m, $R=50$
nm, and $d_{w}=13$ nm is presented in Fig. 5. In this loop we have
subtracted a paramagnetic background. Our results are combined in the lower
panel of Fig. 6. Experimental data for the coercivity of the array are
depicted by dots. In this figure we can observe the strong dependence of the
coercivity as a function of the tube wall thickness, evidencing clearly the
existence of a maximum. Also the coercivities of an isolated tube and
interacting array obtained from our calculations are depicted in the same
figure by dashed and solid lines, respectively. We consider $\varepsilon=20$
in Eq. (4). Note the good agreement between experimental datapoints and
analytical results for interacting arrays for $d_{w}\geq 8$ nm.

\begin{figure}[h]
\begin{center}
\includegraphics[width=6cm, height=6cm]{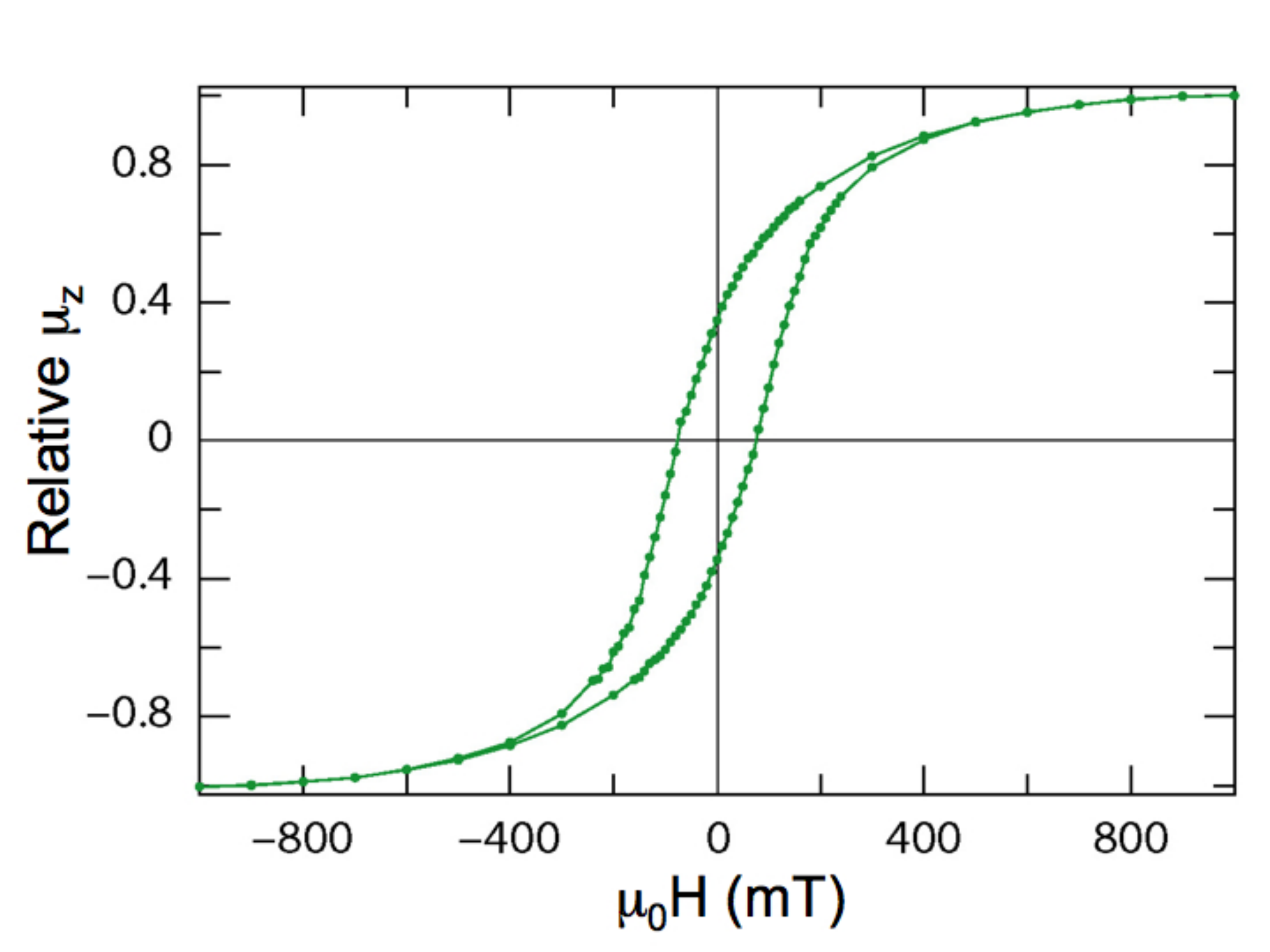}
\end{center}
\caption{(Color online) Magnetic hysteresis loop measured for $L=3$ $\protect%
\mu$m, $R=50$ nm, and $d_{w}=13$ nm (after substraction of a paramagnetic
background and normalization.)}
\end{figure}

\begin{figure}[h]
\begin{center}
\includegraphics[width=7cm]{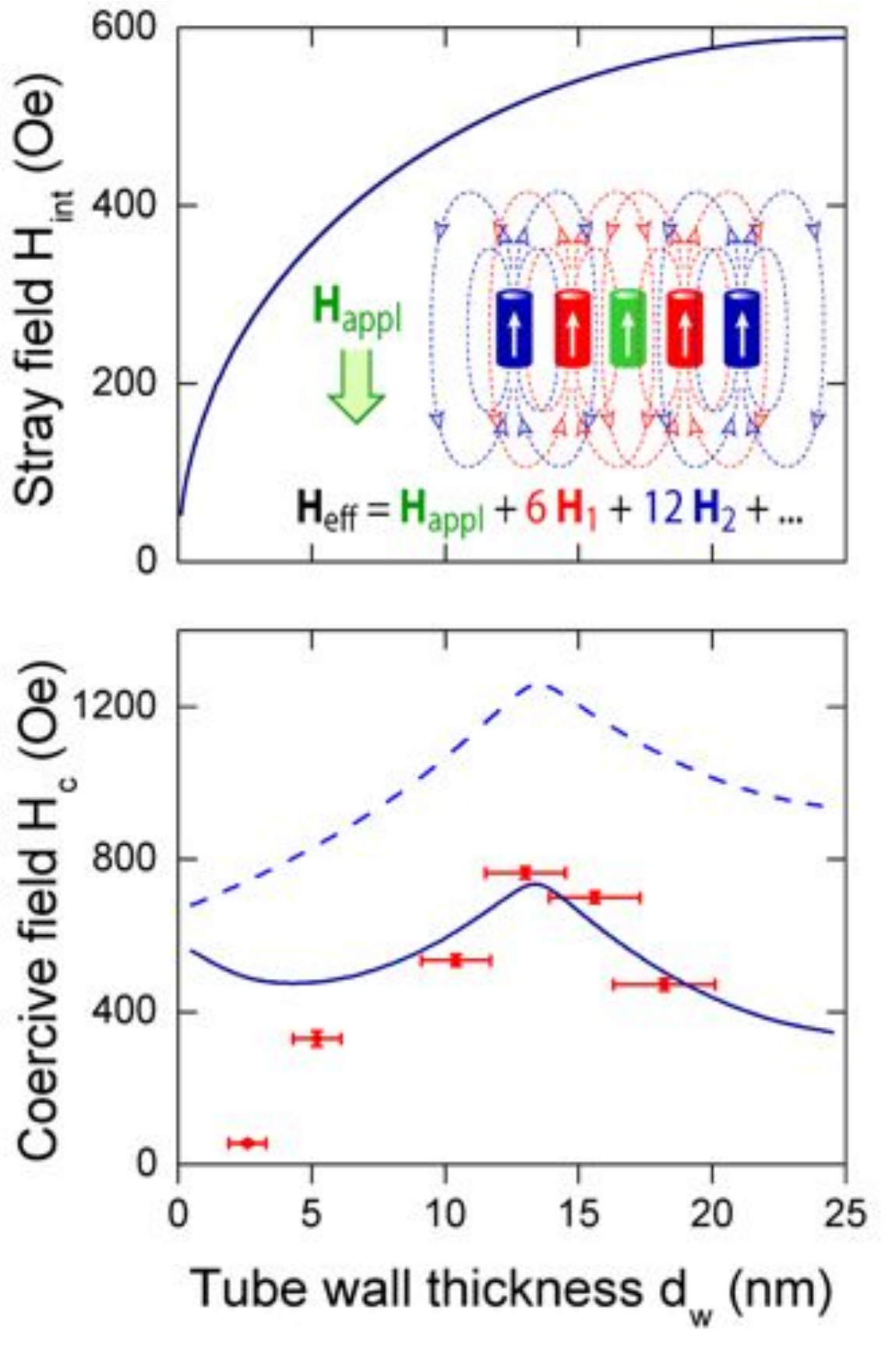}
\end{center}
\caption{(Color online) \textit{(Upper panel)} Stray field $H_{int}$
calculated from Eq. (5) as a function of the tube wall thickness $d_{w}$. As
shown in the inset, in an array, contributions from the six nearest
neighbors, the twelve second-degree neighbors, and the tubes situated
farthest away from the probe tube combine to form the so-called stray field,
which is in the $-z$ direction and therefore drives the probe tube to
reverse its magnetization at smaller absolute values of the applied field
than it would if it were isolated. \textit{(Lower panel)} Coercive field as
a function of the tube wall thickness for Fe$_{3}$O$_{4}$ nanotubes. The red
(dark gray) dots correspond to the data measured in ordered arrays, the
light blue (dashed) line represents the values calculated for an isolated
tube, and the deep blue (solid) curve is calculated for an array of
nanotubes. Parameters: $R=25$ nm, $L=3$ $\protect\mu$m and $D=105$ nm.
Equation (5) was used with $\protect\varepsilon=20$.}
\end{figure}

The deviation of the experimental datapoints from the calculated curve for $%
d_{w}\leq 8$  nm likely originates from the structural imperfections of the
tubes. Deposition of the magnetic material may lead to granular walls at the
initial stages of the growth, whereas further increases in wall thickness
accounted in a smoothening. Other factors not account for in our theoretical
model include thermal instability, which has a stronger effect in thinner
particles, possible shape irregularities, \cite{Braun93} and the finite
length of the tubes.

The results presented above may be generalized. We now proceed to
investigate how the curve will be affected by changes in the tube radius, $R$%
, and of the material, by considering the trajectories of the transition
thickness, $d_{w}^{V-T}$. Such trajectories are shown in Fig. 7 for four
different materials. In the range of parameters considered, we observe that
an increase of the external radius, $R$, results in an increase of the
transition thickness, $d_{w}^{V-T}$. Furthermore, the curves $%
d_{w}^{V-T}\left( R\right) $ are steeper for materials with longer exchange
lengths. Figure 7 can also be interpreted as a phase diagram, in that each
line separates the T mode of magnetization reversal, which prevails in the
upper left region of the ($R$, $d_{w}$) space, from the V mode, found in the
lower right area.

\begin{figure}[h]
\begin{center}
\includegraphics[width=8cm]{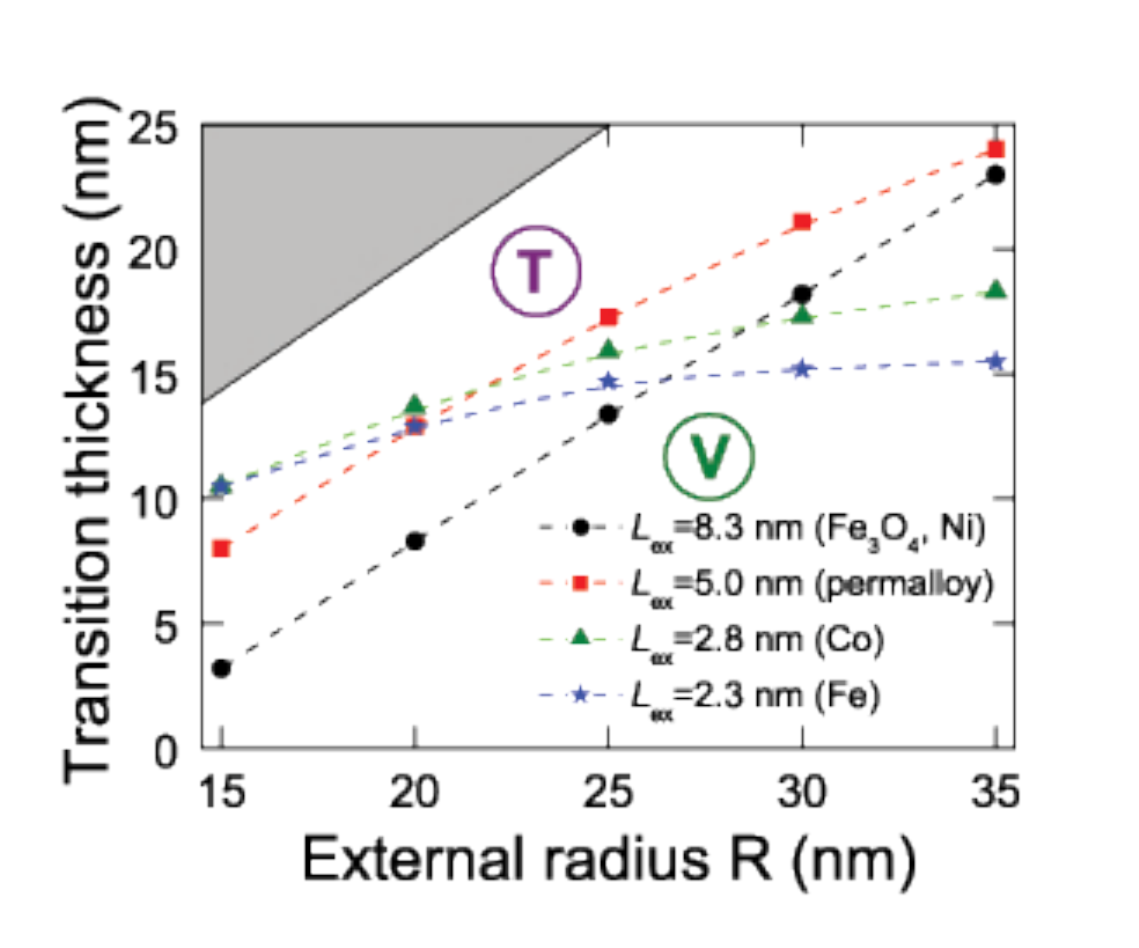}
\end{center}
\caption{(Color online) Trajectories of the transition thickness, $%
d_{w}^{V-T}$, as a function of $R$, for Fe$_{3}$O$_{4}$ (circles, Permalloy
(squares), Co (triangles) and Fe (stars). Because of the similarity of Ni
and Fe$_{3}$O$_{4}$ in terms of their magnetic parameters, the results
presented for Fe$_{3}$O$_{4}$ approximate the case of Ni tubes as well.}
\end{figure}

\section{Conclusions}

In conclusions, by means of simple models for the domain boundary that
appears during the magnetization reversal in nanotubes, we can calculate the
coercive field in ordered arrays of ferromagnetic nanotubes as a function of
the tube wall thickness and the radius. A transition between two different
modes of magnetization reversal, from a vortex boundary, in thin tubes, to a
transverse boundary, in thick tubes, is responsible for the non-monotonic
behavior of the coercivity as a function of wall thickness experimentally
observed. The effect of the stray field originating from the magnetostatic
interactions between the tubes of the array must be included to obtain a
quantitative agreement between experimental and theoretical results. Because
of its long range, the magnetostatic interaction strongly influences the
coercivity of the array. Finally, the presence of a coercivity maximum at a
certain optimum wall thickness should be a quite general phenomenon,
observable for a variety of ferromagnetic materials and of tube radii.
Experimental work remains to be done in order to validate these predictions.

\begin{acknowledgments}
We thank Sanjay Mathur and Sven Barth (Leibnitz Institute of New Materials,
Saarbruecken, Germany) for providing the Fe$_{2}$(O$^{t}$Bu)$_{6}$
precursor. This work was supported by the German Federal Ministry for
Education and Research (BMBF, project 03N8701), Millennium Science Nucleus
Basic and Applied Magnetism (project P06-022F), AFOSR (Award
FA95550-07-1-0040) and Fondecyt (N$^{0}$ 11070010 and 1080300). J. B. acknowledges the Alexander von Humboldt Foundation
for a postdoctoral fellowship (3-SCZ/1122413 STP).
\end{acknowledgments}


\begin{thebibliography}{99}
\bibitem{SMW+00} S. Sun, C. B. Murray, D. Weller, L. Folks, and A. Moser,
Science \textbf{287}, 1989 (2000).

\bibitem{KDA+98} R. H. Koch, J. G. Deak, D. W. Abraham, P. L. Trouilloud, R.
A. Altman, Yu Lu, W. J. Gallagher, R. E. Scheuerlein, K. P. Roche, and S. S.
P. Parkin, Phys. Rev. Lett. \textbf{81}, 4512-4515 (1998).

\bibitem{CKA+99} R. P. Cowburn, D. K. Koltsov, A. O. Adeyeye, M. E. Welland,
and D. M. Tricker, Phys. Rev. Lett. \textbf{83}, 1042-1045 (1999).

\bibitem{WAB+01} S. A. Wolf, D. D. Awschalom, R. A. Buhrman, J. M. Daughton,
S. von Molnar, M. L. Roukes, A. Y. Chtchelkanova, and M. Treger, Science 
\textbf{294}, 1488-1495 (2001).

\bibitem{GBH+02} Th. Gerrits, H. A. M. van den Berg, J. Hohlfeld, L. Bar,
and Th. Rasing, Nature \textbf{418}, 509-512 (2002).

\bibitem{ET03} D. F. Emerich, and C. G. Thanos, Expert Opin. Biol. Ther. 
\textbf{3}, 655-663 (2003).

\bibitem{PKA+01} V. F. Puntes, K. M. Krishnan, and A. P. Alivisatos, Science 
\textbf{291}, 2115 (2001).

\bibitem{WDM+96} W. Wernsdorfer, B. Doudin, D. Mailly, K. Hasselbach, A.
Benoit, J. Meier, J. -Ph. Ansermet, and B. Barbara, Phys. Rev. Lett. \textbf{%
77}, 1873-1876 (1996).

\bibitem{AAX+03} D. Atkinson, A. Allwood, G. Xiong, M. D. Cooke, C. C.
Faulkner, and R. P. Cowburn, Nature Mat. \textbf{2}, 85-87 (2003).

\bibitem{THJ+06} L. Thomas, M. Hayashi, X. Jiang, R. Moriya, C. Retener, and
S. S. P. Parkin, Nature \textbf{443}, 197-200 (2006).

\bibitem{Cowburn07} R. P. Cowburn, Nature \textbf{448}, 544-545 (2007).

\bibitem{Eisenstein05} M. Eisenstein, Nature Methods \textbf{2}, 484 (2005).

\bibitem{CYD+08} Yumei Chen, Fan Yang, Yu Dai, Weiqi Wang, and Shengli Chen,
J. Phys. Chem. C \textbf{112}, 1645-1649 (2008).

\bibitem{Iijima91} S. Iijima, Nature \textbf{354}, 56 (1991).

\bibitem{MLT+02} D. T. Mitchell, S. B. Lee, L. Trofin, N. Li, T. K. Nevanen,
H. Soderlund, and C. R. Martin, J. Am. Chem. Soc. \textbf{124}, 11864-11865
(2002).

\bibitem{SRH+05} S. J. Son, J. Reichel, B. He, M. Schushman, and S. B. Lee,
J. Am. Chem. Soc. \textbf{127}, 7316-7317 (2005).

\bibitem{NCM+05} K. Nielsch, F. J. Casta--o, S. Matthias, W. Lee, and C. A.
Ross, Adv. Eng. Mater. \textbf{7}, 217 (2005).

\bibitem{Wang05} Z. K. Wang, H. S. Lim, H. Y. Liu, S. C. Nq, M. H. Kuok, L.
L. Tay, D. J. Lockwood, M. G. Cottam, K. L. Hobbs, P. R. Larson, J. C. Keay,
G. D. Lian, and M. B. Johnson, Phys. Rev. Lett. \textbf{94}, 137208 (2005).

\bibitem{FMY+06} T. Feifei, G. Mingyun, J. Yuan, Z. Jianmin, X. Zheng, and
X. Ziling, Adv. Mater. \textbf{18}, 2161 (2006).

\bibitem{ELA+07} J. Escrig, P. Landeros, D. Altbir, E. E. Vogel, and P.
Vargas, J. Magn. Magn. Mater. \textbf{308}, 233-237 (2007).

\bibitem{SSS+04} Y. C. Sui, R. Skomski, K. D. Sorge, and D. J. Sellmyer,
Appl. Phys. Lett. \textbf{84}, 1525 (2004).

\bibitem{BJK+07} J. Bachmann, J. Jing, M. Knez, S. Barth, H. Shen, S.
Mathur, U. Gosele, and K. Nielsch, J. Am. Chem. Soc. \textbf{129}, 9554-9555
(2007).

\bibitem{remanence} That is, the \textit{x} and \textit{y} intercepts of the
magnetic hysteresis curve, respectively: the remanence is the fraction of
the saturated (maximum) magnetization that remains once the applied magnetic
field has been turned off, and the coercive field is the absolute value of
the magnetic field applied in the $-z$ direction needed to reduce the
magnetization in the $+z$ direction to zero (in other words, to erase the
remanence).

\bibitem{GSA+05} S. Goolaup, N. Singh, A. O. Adeyeye, V. Ng, and M. B.
Jalil, Eur. Phys. J. B \textbf{44}, 259 (2005).

\bibitem{MF95} H. Masuda, and K. Fukuda, Science \textbf{266}, 1466-1468
(1995).

\bibitem{NCS+02} K. Nielsch, J. Choi, K. Schwim, R. B. Wehrspohn, and U.
Gosele, Nano Lett. \textbf{2}, 677-680 (2002).

\bibitem{Puurunen05} R. L. Puurunen, Appl. Phys. Lett. \textbf{97}, 121301
(2005), and references therein.

\bibitem{LRG03} See for example, B. S. Lim, A. Rahtu, and R. G. Gordon,
Nature Mat. \textbf{2}, 749-754 (2003).

\bibitem{LAE+07} P. Landeros, S. Allende, J. Escrig, E. Salcedo, D. Altbir,
and E. E. Vogel, Appl. Phys. Lett. \textbf{90}, 102501 (2007).

\bibitem{Ohandley00} R. C. O'Handley, Modern Magnetic Materials, Wiley, New
York (2000).

\bibitem{coercivity} Strictly speaking, the determination of the coercivity
actually requires an analysis of the nonlinear regime, which is lacking at
this point. Landeros \textit{et al}. [29] demonstrated that the coherent
mode is present only in very short tubes, and that if there is no other
switching mode than V and T, results for the nucleation field corresponds to
the coercivity.

\bibitem{SW48} E. C. Stoner, and E. P. Wohlfarth, Philos. Trans. R. Soc.
London, Ser. \textbf{A240}, 599 (1948). [Reprinted in IEEE Trans. Magn. 
\textbf{27}, 3475 (1991)].

\bibitem{CLY94} Ching-Ray Chang, C. M. Lee, and Jyh-Shinn Yang, Phys. Rev. B 
\textbf{50}, 6461 (1994).

\bibitem{AS58} A. Aharoni, and S. Shtrikman, Phys. Rev. \textbf{109}, 1522
(1958).

\bibitem{BB92} J. S. Broz, and W. Baltensperger, Phys. Rev. \textbf{45},
7307 (1992).

\bibitem{Brown58} W. F. Brown, Jr., J. Appl. Phys. \textbf{29}, 470 (1958).

\bibitem{Hertel01} R. Hertel, J. Appl. Phys. \textbf{90}, 5752 (2001).

\bibitem{VPH+04} M. Vazquez, K. Pirota, M. Hernandez-Velez, V. M. Prida, D.
Navas, R. Sanz, F. Batallan, and J. Velazquez, J. Appl. Phys. \textbf{95},
6642 (2004).

\bibitem{BAA+06} M. Bahiana, F. S. Amaral, S. Allende, and D. Altbir, Phys.
Rev. B \textbf{74}, 174412 (2006).

\bibitem{ELP+08} J. Escrig, R. Lavin, J. L. Palma, J. C. Denardin, D.
Altbir, A. Cortes, and H. Gomez, Nanotechnology \textbf{19}, 075713 (2008).

\bibitem{Sharrock94} Sharrock M P 1994 \textit{J. Appl. Phys. }\textbf{76}
6413-6418.

\bibitem{ACT+01} Paolo Allia, Marco Coisson, Paola Tiberto, Franco Vinai,
Marcelo Knobel, M. A. Novak, and W. C. Nunes, Phys. Rev. B \textbf{64},
144420 (2001).

\bibitem{Braun93} Hans-Benjamin Braun, Phys. Rev. Lett. \textbf{71}, 3557
(1993).
\end{thebibliography}
\end{document}